\documentclass[twocolumn,american,aps,revmod,superscriptaddress,nofootinbib,showpacs]{revtex4-1}

\usepackage[english]{babel}
\usepackage[T1]{fontenc}
\usepackage{setspace}
\usepackage{outlines}

\usepackage[colorinlistoftodos]{todonotes}
\usepackage[colorlinks=true, allcolors=blue]{hyperref}
\usepackage{amsmath}
\usepackage{xcolor}
\usepackage{placeins}
\usepackage{booktabs}
\usepackage{float}
\usepackage{graphicx}

\makeatletter
\renewcommand{\l@section}{\@dottedtocline{1}{1.em}{1.9em}}
\renewcommand{\l@subsection}{\@dottedtocline{2}{2.0em}{1.5em}}
\renewcommand{\l@subsubsection}{\@dottedtocline{3}{3.em}{1.5em}}
\makeatother

\begin{document}

\title{Numerical Investigation of Non-equilibrium Electron Effects on the Collisional Ionization Rate in the Collisional-Radiative Model}
\author{M. S. Cho}
\affiliation{Gwangju Institute of Science and Technology, Department of Physics and Photon Science, Gwangju 61005, South Korea}
\affiliation{Lawrence Livermore National Laboratory, 7000 East Avenue, Livermore, California 94550, USA}
\author{H. -K. Chung}
\affiliation{Korea Institute of Fusion Energy, Daejeon 34133, South Korea}
\author{M. E. Foord}
\affiliation{Lawrence Livermore National Laboratory, 7000 East Avenue, Livermore, California 94550, USA}
\author{S. B. Libby}
\affiliation{Lawrence Livermore National Laboratory, 7000 East Avenue, Livermore, California 94550, USA}
\author{B. I. Cho}
\email{bicho@gist.ac.kr}
\affiliation{Gwangju Institute of Science and Technology, Department of Physics and Photon Science, Gwangju 61005, South Korea}

\date{\today}
\begin{abstract}
The interplay of kinetic electron physics and atomic processes in ultrashort laser-plasma interactions provides a comprehensive understanding of electron energy distribution's impact on plasma properties. Notably, non-equilibrium electrons play a vital role in collisional ionization, influencing ionization degrees and spectra. This paper introduces a computational model that integrates the physics of kinetic electrons and atomic processes, utilizing a Boltzmann equation for non-equilibrium electrons and a collisional-radiative model for atomic state populations. The model is used to investigate the influence of non-equilibrium electrons on collisional ionization rates and their effect on population distribution, as demonstrated by L. Young \cite{young2010femtosecond}. The study reveals significant non-equilibrium electron presence during XFEL-matter interactions, profoundly affecting collisional ionization rates in the gas plasma, thereby necessitating careful consideration of the Collisional-Radiative (CR) model applied to such systems.

\end{abstract}

\maketitle
\section{Introduction}
Recently, the examination of the interplay between kinetic physics of electrons and atomic processes in the ultrashort region of laser-plasma has become possible through the development of ultrafast laser technology \cite{ditmire1998time, hansen2002hot,le2019influence,medvedev2011short,milder2021measurements, lee2021investigation, bang2022review}. This advance has allowed for a direct and more comprehensive analysis of the electron energy distribution and its impact on plasma properties, beyond the traditional approach of solely measuring the temperature and density of free electrons. The electron distribution undergoes a temperature process until it reaches a specific distribution, with the presence of non-equilibrium electrons during this process having a substantial effect on the characteristics of plasmas. This necessitates a new interpretation of existing plasma phenomena, with non-equilibrium electrons playing a crucial role in various laser heating processes, such as inverse bremsstrahlung \cite{milder2021measurements} and thermalization \cite{lee2021investigation}. These findings have significant implications for the development of plasma-based technologies and applications.

In this context, researchers also have been exploring the impact of non-equilibrium electrons on plasma properties in the Collisional-Radiative (CR) model, commonly used to calculate the ionization degree and spectrum. When calculating the collisional ionization rate in the CR model, the electron energy distribution function of free electrons is required, and it is typically assumed to follow a specific statistical distribution, such as Maxwell-Boltzmann or Fermi-Dirac distribution, for calculation-costly efficiency. While this is a reasonable assumption for high-density plasmas with short thermalization times, the use of targets with various materials and densities necessitates a precise understanding of the effects of non-equilibrium electrons in the ultrashort region. To address this need, researchers have attempted to incorporate the effect of non-equilibrium electrons in the CR model \cite{le2019influence, de2013non,ren2023simulations,abdallah2013time,gao2015evolution}.

SCFLY, one of widely used CR codes, has been successful in revealing the main ion dynamics occurring in X-ray Free Electron Laser (XFEL) heated plasmas and modeling their consequential transmission/emission spectrum \cite{ciricosta2011simulations,vinko2012creation,cho2012resonant,ciricosta2012direct,ciricosta2016measurements,cho2017observation,cho2018intensity,van2018clocking}; however, most of the calculations assume a Maxwell-Boltzmann distribution for the free electron energy, which may not accurately reflect the behavior of low-density targets such as gas targets. An experiment using an XFEL to heat a low density neon gas target is one of examples to use SCFLY for interpretation \cite{ciricosta2011simulations}, but the code did not sufficiently consider the effect of non-equilibrium electrons, necessitating code improvements to ensure accurate analysis of low-density plasma dynamics. 

This paper presents a model for calculating non-equilibrium electron distributions coupled with the code SCFLY, which is used to reinterpret an XFEL heated neon plasma experiment which has reported the difference between experiment and theory \cite{young2010femtosecond}. The result of this work reveals existence of non-equilibrium plasma electrons, confirming that their impact on the final ion distribution is not negligible. Here, we find that by including the calculated non-Maxwellian distributions in the atomic modeling, much better agreement is found for ionization balance of the Neon plasmas. The module is developed to solve the Boltzmann transport equation under the assumption of spatially unform plasmas in cooperation with 0-d Collisional-Radiative philosophy and updating electron distribution function in time. Its impact on the transition rate, e.g. the collisional ionization rate, results in the change of plasma characteristics such as the charge state distribution, which closely approaches the experimental data. In addition, it also probes the differences between different collision-ionization coefficient models and confirms that there is no significant difference in the corresponding experimental density. These findings demonstrate the accuracy and broad applicability of SCFLY in interpreting low-density plasma dynamics, improving our understanding of complex plasma behavior.

\section{Computational Model}
\subsection{Collisional-Radiative Model}
The CR model provides population distributions and characterizes the physical processes occurring in the plasma for a given electron temperature and density. It can be considered as the most general population kinetics calculational model since it uses the fundamental physical quantities, e.g. transition rate of state kinetics to get population instead of statistical, local thermodynamic equilibrium (LTE). CR model solves sets of rate equations to calculate the number density of the $i^{th}$ atomic state $N_i$ as a function of rates $A_{j \rightarrow i}$ from the jth atomic state to the $i^{th}$ atomic state, where 1 $\le$ i, j $\le$ m (the maximum number of atomic states) from the following equation.

\begin{equation} \label{rate_eq}
    \frac{dN_{i}}{dt} = \sum_{j \neq i}A_{j \rightarrow i}N_{j} - A_{i}^{*}N_{i}
\end{equation}
where 
\begin{equation*}
    A_{i}^{*} = \sum_{k \neq i} A_{i\rightarrow k} 
\end{equation*}

For this work, the code uses the super-configuration model with the screened hydrogenic levels whose energy level is based on principal quantum numbers considering the screened nuclear charges seen by the electrons in each shell. This is a linear function of the shell occupations using a set of screening coefficients. The transition rate $A_{i->j}$ is the transition rate from level $i$ to level $j$. Among them, the rate of a collisional transition $A^{COL}$ includes the collisional cross-section term $\sigma(E)$ as a fundamental physical quantity and, thus, the collisional rate is obtained by integrating over the electron energy distribution function $f_{e}(E)$ as

\begin{equation} \label{rate_c}
   A^{COL} = n_{e} \int_{\Delta E}^{\infty} v \sigma(E) f_{e}(E)\,dE
\end{equation}

where \(n_e\) is the electron density, \(v\) is the electron velocity at energy \(E\), and \(\Delta E\) represents the threshold energy for the transition. The electron energy distribution $f_{e}(E)$ in Eq. \ref{rate_c} has been assumed to be the Maxwell-Boltzmann distribution which implies ionized-electrons thermalized instantaneously in the code SCFLY. This work supplements that module with the Boltzmann equation solver for non-Maxwellian electrons evolution in time and updates $f_{e}(E)$ to reveal the effect of it on the total ionization in XFEL-heated plasmas. The Boltzmann equation solver is described in section B in detail. 

Another important term in the collisional rate is the collisional cross-section $\sigma(E)$ derived as a few different models for plasmas. For the collisional ionization and recombination process, the collisional ionization cross-section in the code is used as a semi-empirical formula in the code. Burgess-Chidichimo (BC) models have been generally used as a basic option whose Collisional Ionization (CI) cross-section is given by

\begin{equation} \label{ci_lotz}
    \sigma^{BC}(E) = \pi a_{0}^{2} C \xi (\frac{I_{H}}{\Delta I})^{2} (\frac{\Delta I}{E}) \log (\frac{E}{\Delta I}) W(\frac{E}{\Delta I})
\end{equation}

where 

\begin{equation*} 
    W(\frac{E}{\Delta I}) = [\log (\frac{E}{\Delta I})]^{\frac{\beta \Delta I}{E}}
\end{equation*}
and 
\begin{equation*} 
    \beta = 0.25((\frac{100z+91}{4z+3})^{1/2}-5)
\end{equation*}

When the plasma is ionized from an initial level \(I\) for an electron with energy \(E > \Delta I\) (where \(\Delta I\) is the ionization potential) \cite{burgess1983electron}, \(z\) represents the charge of the ion and \(\xi\) is the effective number of equivalent electrons in a shell. $a_{0}$ represents the Bohr radius, the constant \(C\) is often assigned a suggested value of 2 \cite{chung2005flychk}, and $I_{H}$ is Rydberg constant in the formula. There are several CI cross-section models available, such as the Lotz model \cite{lotz1968electron}, the BCF model \cite{van2018clocking}, and others. It is known that these models yield similar values in the density region of approximately \(10^{19} cm^{-3}\). This indicates that the result is insensitive to the choice of the collisional cross-section model \cite{van2018clocking}. Thus, we have chosen to use the BC model for this work. 

For the three-body recombination rate $A^{RC}$ with an arbitrary electron distribution, one must consider the differential collisional recombination (DCR) cross-section, denoted as $\sigma^{DCR}(f \rightarrow i)(E', E'' \rightarrow E)$. This cross-section is derived from the detailed balance with the differential collisional ionization cross-section, known as the Fowler relation \cite{oxenius1986kinetic, ralchenko2016modern}. Thus, the three-body recombination rate using the DCR cross-section can be expressed as:

\begin{multline*} 
A^{RC} = n_e^2 \int \int \int \left( \frac{2E'}{m_e} \right)^{\frac{1}{2}} \left( \frac{2E''}{m_e} \right)^{\frac{1}{2}} f_e(E') f_e(E'') \\
\sigma^{DCR}(f \rightarrow i)(E', E'' \rightarrow E) dE' dE'' dE
\end{multline*} 

where $m_e$ is the electron mass, and the energy $E$ refers to incoming electron energy, while the energies of the ejected and outgoing electrons are represented as $E'$ and $E''$, respectively.

Auger ionization and its inverse process, dielectronic recombination, are pivotal phenomena in plasma environments. The CR model approximates the rates of these processes using a detailed balance approach. The Auger ionization rate \(A_{\text{aug}}(k \rightarrow i)\), which describes the transition from state \(k\)—an excited state of an ion plus its outermost excited electron—to a bound state \(i\) of the subsequent ion, is given by

\begin{multline*} 
\frac{n_e \int \sigma^{EC}(i \rightarrow k) v f_e(E) dE}{A^{aug}(k \rightarrow i)} = \\
\frac{g_k}{g_i} \frac{n_e}{2} \left( \frac{h^2}{2 \pi m_e k_B T_e} \right)^{3/2} e^{-\frac{(E_k-E_i)}{k_B T_e}}
\end{multline*} 

where \(g_k\) and \(g_i\) denote the degeneracy factors of states \(k\) and \(i\), respectively, \(h\) is the Planck constant, \(k_B\) is the Boltzmann constant, and \(T_e\) represents the electron temperature. The term \(\sigma^{\text{EC}}(i \rightarrow k)\) refers to the electron capture cross-section from state \(i\) to \(k\), and \(f_e(E)\) is the electron energy distribution function (EEDF), indicating the influence of free electron distribution on Auger ionization and its reverse process. This equation highlights the intricate relationship between electron dynamics and ionization mechanisms within plasmas, based on the principle of detailed balance. Detailed explanation for the theory of CR model and the code SCFLY itself is on the ref.\cite{chung2005flychk,chung2007extension,ralchenko2016modern,chung2017atomic,cho2020opacity,sohn2023opacity}.

\subsection{Boltzmann Solver under the Rate Formalism}

The time evolution of the electron distribution function used for the collisional rate calculation is calculated from the well-known kinetic model, the Boltzmann transport equation. The general form of the Boltzmann transport equation is given by 

\begin{multline} \label{transport}
(\frac{\partial}{\partial t} + v \cdot \nabla_{r} + \frac{eE}{m} \cdot \nabla_{r})f(r,v,t) =\\ 
C_{ee}^{elas}(f) + C_{ei}^{elas}(f) + Q^{inelas}(f) 
\end{multline}

where the function $f(r,v,t)$ is the distribution function for electrons at time $t$, and spatial location $r$ with velocity $v$. For the consistency with the zero dimensionality in the CR code, it is assumed that the electron energy distribution f is spatially uniform. Also, the velocity term $v$ in the distribution can be simplified with a two-term spherical harmonic expansion \cite{morgan1990elendif} and the effect of the external electric field $E$ on the change of the electron distribution is assumed to be minor in XFEL heated plasmas. It results in the second and third terms in the left-hand side turn out to be zero. The term $C^{elas}_{ee}$ represents electron-electron elastic collisions, which account for changes in the distribution resulting from the collisional processes of free electrons themselves, while preserving the total energy of the free-electron system. The term $C^{elas}_{ei}$ describes electron-ion elastic collisions, acting as a momentum transfer operator. During these collisions, electrons transfer energy to ions. Finally, the term \(Q_{ei}^{inelas}\) signifies the alteration of the distribution function due to inelastic collisions, covering atomic kinetic processes such as collisional ionization/three-body recombination, and autoionization/electron capture. These processes either contribute to or deduct from the electron distribution function. Additionally, the model incorporates excitation and de-excitation events associated with free electrons, inducing shifts within the energy spectrum that correspond with the transition energies of bound electrons in ions, thus ensuring energy conservation.

Accordingly, Eq. \ref{transport} can be reformulated into a rate equation with these considerations, defined by the electron number density \(n(\epsilon)\):

\begin{equation} \label{rateeq_ee}
    \frac{\partial n_{k}^{e}}{\partial t} = -\frac{\partial J_{ee}}{\partial \epsilon} - \frac{\partial J_{ei}}{\partial \epsilon} + S(f) + I(f)  
\end{equation}

Here, \(J_{ee}\) and \(J_{ei}\) represent the electron flux along the energy axis, influenced by electron-electron and electron-ion collisions, respectively. The terms \(S(f)\) and \(I(f)\) serve as source and sink terms, respectively, for inelastic processes involving ionization, excitation and their inverse processes, effectively replacing \(Q^{inelas}(f)\) in Eq.~\ref{transport}. These two terms, representing the number of electrons added and subtracted from the EEDF, are derived from atomic transitions at that particular time step, as determined by the rate equation (Eq.~\ref{rate_eq}). Consequently, \(S(f)\) and \(I(f)\) highlight the discrete, stepwise nature of these changes, while the descriptions of \(J_{ee}\) and \(J_{ei}\) capture the continuous evolution of the electron distribution.

Note that this model does not account for the distribution changes directly induced by photons, given that the driving photon energy falls within the x-ray regime, which corresponds to a significantly lower cross-section. Furthermore, the scope of this work does not require the consideration of the self-generated radiation field from plasma, attributed to the system's brief timescale. However, should the driving photon be derived from an optical laser, or should the system operate over a longer timescale ($>$ a few hundred picoseconds), incorporating this aspect into the model becomes essential for accurate analysis.

Specifically, the first term describes the impact of electron-electron (ee) collisions on the distribution function. The work of Rosenbluth et al. demonstrates that the fundamental two-body force, described by an inverse square law, can be formulated as a simple potential term in the Fokker-Planck equation for the distribution \cite{rosenbluth1957fokker, morgan1990elendif}. If the continuous function $n^e_k(\epsilon)$ is put in finite difference form as $n(\epsilon_i)$, the flux can be discretized by projecting it onto a finite-difference energy axis. The derivative of $J_{ee}$ with respect to $\epsilon$ turns out to be a transition rate from $m^{th}$ to $k^{th}$ energy grids, $R_{ee}^{km}$, times the population $n_m$ at the $m^{th}$ energy grid. If one utilizes a time step that is small enough to assume transitions take place between adjacent energy grids, then this term results in the following expression:

\begin{equation} \label{continue}
    R_{km}^{ee}n_{m} \sim I_{k-1}n_{k-1} + O_{k+1}n_{k+1} - (I_{k}+O_{k})n_{k} 
\end{equation}
where the rate coefficients I and O are
\begin{align*} 
    I_{k} = \sum_{j}w_{kj}n_{j},  O_{k} = \sum_{k}w_{kj}n_{k}
\end{align*}
and
\begin{align*} 
    w_{kj} = [w_{kj}^{'}w_{j-1,k+1}^{'}(\epsilon_{j-1}\epsilon_{k+1})^{1/2}(\epsilon_{j}\epsilon_{k})^{-1/2}]^{1/2}
\end{align*}
whose $w_{kj}^{'}$ is given by

{\small\begin{align*}
  w_{kj}^{'} =
    \begin{cases}
      \alpha [\epsilon_{k+1}^{-1/2}+\epsilon_{k}^{-1/2}] (\epsilon_{j}\epsilon_{k}-\frac{3}{4}) & \text{ $k > j$ }\\
      \alpha [\epsilon_{k+1}+\epsilon_{k}] u_{k}\epsilon_{j}^{-1/2} & \text{ $k < j-1$ }\\
      \alpha [\epsilon_{k+1}^{-1/2}+\epsilon_{k}^{-1/2}] (\epsilon_{j}\epsilon_{k}-\frac{3}{4}) + \epsilon_{k}u_{k}\epsilon_{j}^{-1/2} & \text{ $k = j$ }\\
      \alpha [(\epsilon_{j}u_{k}-\frac{3}{4})\epsilon_{k+1}^{-1/2}+(\epsilon_{k+1}+\epsilon_{k})u_{k}\epsilon_{j}^{-1/2}] & \text{ $k = j - 1$ }\\
      0 & \text{ $k = 1$ or $k_{max}$ }\\
    \end{cases}       
\end{align*}}

Here, $\alpha$ a coefficient that encapsulates the information regarding the collisional frequency between electrons and given by
\begin{equation} \label{coeff1}
   \alpha = \frac{2}{3}\pi e^{4} (\frac{2}{m})^{1/2} \ln{\Lambda}
\end{equation}

and $e$ is the charge of electron, $m$ is the electron mass and $\ln \Lambda$ is the coulomb logarithm. The detailed derivation from the continuous equation is shown in the reference \cite{rockwood1973elastic, elliott1976electron, morgan1990elendif}. For reference, the code uses the Coulomb logarithm derived from the Spitzer formula for electrons with lower energy than the plasma frequency, and it employs an empirical formula, based on the work of Zollweg \& Liebermann et. al. \cite{zollweg1987electrical}, for other cases. This is necessary because the Spitzer formula may not yield accurate results for non-ideal plasmas, particularly when the electron kinetic energy approaches the Coulomb potential. In such scenarios, only a few electrons, or even a fraction of an electron, occupy a Debye sphere and this phenomenon typically occurs at relatively low temperatures but relatively high electron densities.

The second term of Eq. \ref{rateeq_ee} is the momentum transfer operator, calculating the energy loss of electrons due to the elastic collision with the ions. Similar with the electron-electron collision term, it can be converted to the finite difference form with the rate equation form as 

\begin{equation} \label{continue2}
    R_{km}^{ei}n_{m} \sim r_{k-1}n_{k-1} + d_{k+1}n_{k+1} - (r_{k}+d_{k})n_{k} 
\end{equation}

and $r_{k}$ and $d_{k+1}$ are given by;

\begin{equation*} \label{ak1}
    r_{k} = \frac{\bar{\nu_{k}}}{2 \Delta \epsilon} (\frac{kT_{i}}{2} - \epsilon^{+}_{k} + \frac{2kT_{i}}{\Delta \epsilon} \epsilon^{+}_{k})
\end{equation*}

\begin{equation*} \label{bkp1}
    d_{k+1} = \frac{\bar{\nu_{k}}}{2 \Delta \epsilon} (-\frac{kT_{i}}{2} + \epsilon^{+}_{k} + \frac{2kT_{i}}{\Delta \epsilon} \epsilon^{+}_{k})
\end{equation*}

where $\epsilon_{k}^{+}=k_{B} \Delta \epsilon$, $\frac{\nu^{+}_{k}}{N} = (\frac{2 \epsilon^{+}_{k}}{m})^{1/2} \sum_{s}q_{s}\sigma_{s}(\epsilon^{+}_{k})$, and $ \bar{\nu_{k}^{+}} = 2mN(\frac{2\epsilon^{+}_{k}}{m})^{1/2} \sigma_{s}(\frac{q_{s}\sigma_{s}(\epsilon^{+}_{k})}{M_{s}})$ \cite{rockwood1973elastic}. The matrix $R$, having dimensions of $s^{-1}$, represents the collision frequencies for the flow of electrons up and down the energy axis due to elastic collisions with ions. The excitation rate $r_{k}$ corresponds to the excitation rate of electrons in the electron energy distribution, and $d_{k}$ corresponds to the de-excitation rate of electrons. The constant $k_{B}$ is the Boltzmann constant, and $T_i$ is the ion temperature. One of the key assumptions for converting the electron flux into a discretized form for both electron-electron and electron-ion elastic collisions is the use of a short calculation timestep. This timestep must be short enough to allow the majority of the electron population at energy \(k\) to move to adjacent energy bins. As a result, the R-matrix should be diagonally dominant.

Also, this formalism requires the temporal development of the distribution function that will be represented accurately only for “small” $\delta t$. The source $S(f)$ and sink terms $I(f)$ in the Eq. \ref{rateeq_ee} describe changes in the number of electrons associated with bound-free transitions in ions. These transitions occur due to the ionization and recombination processes resulting from electron collisions with ions. The ionization or recombination process involves the conversion of bound electrons to free electrons and vice versa, which cannot be solved using the rate equation formalism alone. To address this challenge, the source and sink terms are incorporated after solving the rate equation. By utilizing the ionization (or recombination) information calculated within the rate equation of the collisional-radiative model, it becomes possible to determine the number and energy of electrons emitted during the transition. This information is then used to formulate the corresponding source (or sink) term. Note that using a very small-time step is necessary to prevent divergence. Increasing the time step leads to a significant amount of addition or subtraction occurring at once. If the value to be added exceeds the density of electrons' states at the corresponding energy or the value to be subtracted becomes smaller than 0, the calculation will halt.

\section{Numerical Study on the Impact of Non-equilibrium Electrons}
\subsection{Time Evolution of EEDF}

\begin{figure*}
    \centering
    \includegraphics[width=.95\linewidth]{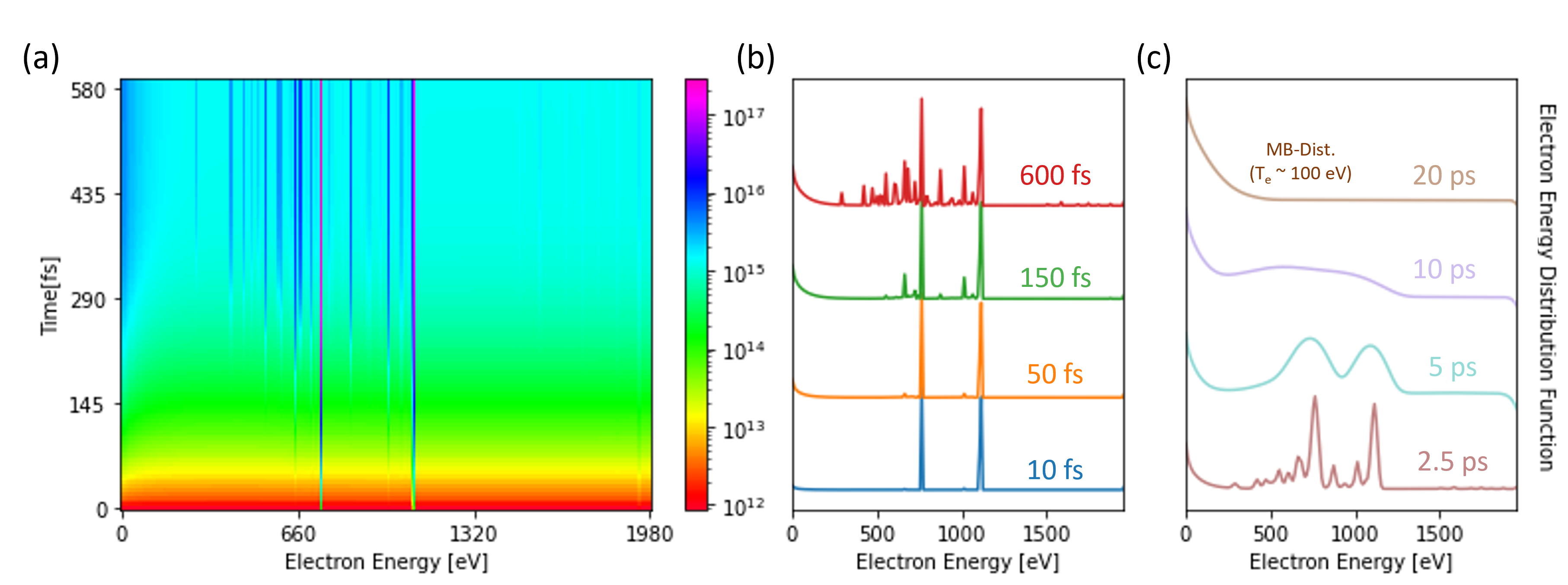}
    \caption{\label{evolution} (a) The calculated electron energy distribution function in $cm^{-3}eV^{-1}$ as a function of electron energy ($eV$) in time during irradiation by 2.5x$10^{18}$ $W/cm^{2}$ XFEL Laser pulse with the pulse duration of 230 $fs$. The pulse center is in 300 $fs$ and the density of neon plasma is 2x$10^{19}$ $cm^{-3}$. (b) The representative electron energy distribution for four different time steps. Two major peaks from KLL Auger decay ($\sim$770 $eV$) and photoionization ($\sim$1130 $eV$) process are observed first early on, followed by other satellite peaks associated with those processes. (c) The evolution of EEDF in longer time scale. Eventually it ends up with the Maxwell-Boltzmann (MB) distribution with Te $\sim$ 100 $eV$ which is the same value estimated by the population kinetic calculation part. Modeling beyond the collisional-radiative model, such as hydrodynamic expansion, conductive energy loss, etc., must be incorporated to accurately describe EEDF in this region, so simulations are conducted to test whether the approach is solely limited to equilibrium conditions.}
\end{figure*}

These methods were applied to the study of ultrafast electron response of atoms to the intense X-ray with Ne plasma heated by XFEL, which was conducted by L. Young, et. al.\cite{young2010femtosecond}. The Neon gas target has low density enough to get slow thermalization of free electrons compared to the pulse duration of XFEL, resulting in the accumulation of the effect of hot electrons with a long lifetime in the plasmas rather than the solid target. Also, the major measurement of the experiment is the charge state distribution (CSD) which can be simulated by the new code in this work.  Young, et. al., illustrates three different systems with three XFEL photon energy cases (800, 1050, 2000 $eV$) representing the different ionization mechanisms, but this work focuses on the case of the XFEL photon with the energy of 2000 $eV$  since (1) all collisional ionization process of both L-shell and K-shell electrons in all charge states, even including H- and He-like ion of Ne, is possible to be analyzed in detail, and (2) the other non-collisional atomic processes, e.g. direct multi-photon process and resonance absorption, are possible to be neglected in the analysis \cite{doumy2011nonlinear,xiang2012inner}. 

In order to check the effect of hot electrons, we simulate the electron energy distribution function in time. In these simulations, several variables affect the results, such as x-ray pulse duration, X-ray intensity, and gas density, but their exact values are unknown. Thus, we based the simulation on commonly used values from previous studies. The previous numerical studies have used a single density of $1.6$ x $10^{19} cm^{-3}$ for neon gas corresponding to a gas pressure of 500 $torr$ used in the experiment \cite{de2013non,abdallah2013time,gao2015evolution}. This gas has been heated up with a XFEL pulse with a Gaussian temporal profile and whose pulse duration is set to be 230 $fs$ reported in the ref.\cite{young2010femtosecond}.  is used in the simulation and the temporal spikes of the intensity profile due to Self-Amplified Spontaneous Emission (SASE) process can be ignored as its effect is expected to be negligible \cite{ciricosta2011simulations,bonifacio1994spectrum}.

The absorption of a 2000 $eV$ XFEL photon by the neon plasma results in changes in the EEDF, as demonstrated in Figure \ref{evolution}. The Boltzmann solver, the new calculation module, tracks the evolution of electrons in response to the ionization processes initiated by the XFEL-plasma interaction. At the start of the simulation, the target is a neutral gas plasma, so the EEDF exhibits a flat shape. Upon XFEL exposure to the neon plasma, ionization begins, with photoionization as the dominant mechanism. During this process, electrons in the K-shell absorb photons and are ionized to be free electrons with energy of approximately 1020 $eV$. The KLL Auger process then fills the resulting K-shell hole with electrons, ionized at an energy of approximately 800 $eV$ (K-shell binding energy $\sim$ 980 $eV$ and L-shell binding energy $\sim$ 90 $eV$). The simulation accurately captures the EEDF at early time steps, showing sharp peaks at the corresponding energies for the neutral neon plasmas. As time goes on, higher charge ions are produced through photoionization and Auger decay, resulting in satellite peaks of lower energy than the initial photo-ionization and Auger electron peaks. These satellite peaks can be seen as multiple electron peaks observed in the EEDFs after 150 $fs$. 

The simulation results reveal that the commonly assumed instant thermalization in the collision-radiative model is not valid within the low-density environment of this experiment. While the collisional frequency in normal solid density targets allows the electron distribution to reach equilibrium within a few femto-seconds, caution is necessary when dealing with femto-second dynamics in low-density targets like gas targets. In this specific case, the simulation demonstrates the presence of energetic non-equilibrium electrons generated by photo-ionization and Auger decay, persisting for several hundred femto-seconds during the interaction between the X-ray pulse and the target. These electrons possess higher energy levels than thermal electrons and constitute a substantial portion of the population, significantly influencing the calculation of the collisional rate. Consequently, for accurate descriptions of such interactions, it is imperative to go beyond the CR model and consider non-equilibrium effects.

The creation of a thermalized electron population in EEDF is mainly attributed to electron-ion and electron-electron collision processes. Upon incidence of the XFEL on neon plasma, photoionization and Auger decay processes are identified as the primary causes of EEDF change, followed by collision processes that result in peak broadening and electron population accumulation at low energy. The EEDF ultimately converges to the MB distribution with $T_{e}$ $\sim$ 100 $eV$ after approximately 20 $ps$, indicating the thermalization time of neon plasma at the density in the simulation. It is based on the collisional frequency formula derived from the Spitzer model \cite{spitzer1953transport} mainly contributing to this density regime. Also, note that the calculation later ($\sim$ pico-second regime) is only considered as the verification of sanity check of the simulation. This code has the nature of zero-dimensional calculation that does not consider spatial relevant physics, such as hydrodynamic expansion of target, spatial dissipation of heat, etc, which may affect the evolution of EEDF in the longer time scale. Nevertheless, the Boltzmann solver effectively depicts the evolution of free electron distribution in XFEL-heated plasma over time and represents an advancement over the instantaneous thermalization assumption employed by existing CR models.

\subsection{Population Changes due to Non-Equilibrium Electrons}

Taking into account the influence of non-equilibrium electrons, the fractional yield of each charge state undergoes changes. A previous study by O. Ciricosta et. al. utilized SCFLY to simulate the charge state yield in detail \cite{ciricosta2011simulations}. In order to compare these results with L. Young's experimental findings, which measured the fractional yield of the charge state, simulations were conducted while considering the XFEL's intensity beam profile. The simulation assumed an elliptical-Gaussian shape for the XFEL intensity beam profile, with axes in the ratio of 1:2 \cite{ciricosta2011simulations,young2010femtosecond}. Moreover, to facilitate comparison with the experimental ion population over time, the fractional yield was determined for the charged states, excluding neutral ions, based on the initial ion density condition.

\begin{figure}
  \centering
  \includegraphics[width=.95\linewidth]{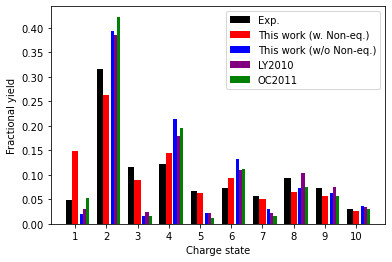}
  \caption{\label{pop_comp} Experimental and simulated charge state populations for Neon plasma heated by a 2000 $eV$ XFEL pulse. The experimental data is depicted in the black bar \cite{young2010femtosecond}, while the simulation results are represented by the various color bars. Notably, the red bar corresponds to a simulation that includes non-equilibrium electrons, while the blue bar reflects the same calculation assuming instant thermalization of free electrons. This instant thermalization scenario closely aligns with the theoretical estimations presented by L. Young et. al. (purple) \cite{young2010femtosecond} and O. Ciricosta et. al. (green) \cite{ciricosta2011simulations} as expected.}
\end{figure}

Figure \ref{pop_comp} represents the results obtained from experimental and calculated fractional yield of the charge state. The distribution is an integrated value of each charge state for the entire time (up to $20$ $ns$ in this work), and has been performed in the same way as the previous works. Detailed calculation methods are introduced in Ref. \cite{ciricosta2011simulations}. Based on Fig. 2, the yields of the +3, +5, +7 charged ions show an evident difference between the experiment and simulations with the instant thermalization assumption, aligning with findings by L. Young and O. Ciricosta. Further research identified various issues, such as the double auger process and cross-section, as contributing factors, supported by calculation \cite{doumy2011nonlinear,xiang2012inner}. However, accounting for non-equilibrium electrons resulted in the fractional yield closely matching the experimental data for populations +2 to +7, especially for the charged ions of +3, +5, and +7. The significance of collisional ionization due to non-equilibrium electrons has not been emphasized in previous studies, possibly due to their low collisional ionization rates stemming from the typically low density \cite{son2011impact}. Nonetheless, our results demonstrate that considering a substantial number of non-equilibrium electrons, as opposed to relying solely on the Maxwell-Boltzmann distribution, can significantly impact the final outcomes. 

Nevertheless, this simulation diverged significantly from the experimental values previously well matched for the +1 ion population. This discrepancy is attributed to the assumption that the electron distribution follows the Maxwell-Boltzmann distribution, which tends to concentrate more electrons in lower energy regions. This leads to increased collisional ionization and a greater population of lower charge states. It is a phenomenon that has been observed to be consistent in other cases, such as with 800 eV. Therefore, to improve the charge state distribution, it would be necessary to extend the simulation time until sufficient generation of low-energy electrons occurs or thermalization takes place. However, such refinement must take into account the effects of hydrodynamic expansion, radiative, and conductive losses in the region, demanding a code capable of incorporating this dimensional information while calculating the non-equilibrium electron distribution. Moreover, the current code needs to account for simulating the double Auger process and cross-section, which previous research has identified as problematic areas \cite{santra2020interaction}.

\subsection{Collisional Ionization Rate Enhanced by Non-equilibrium Electrons }

\begin{figure*}
    \centering
    \includegraphics[width=.95\linewidth]{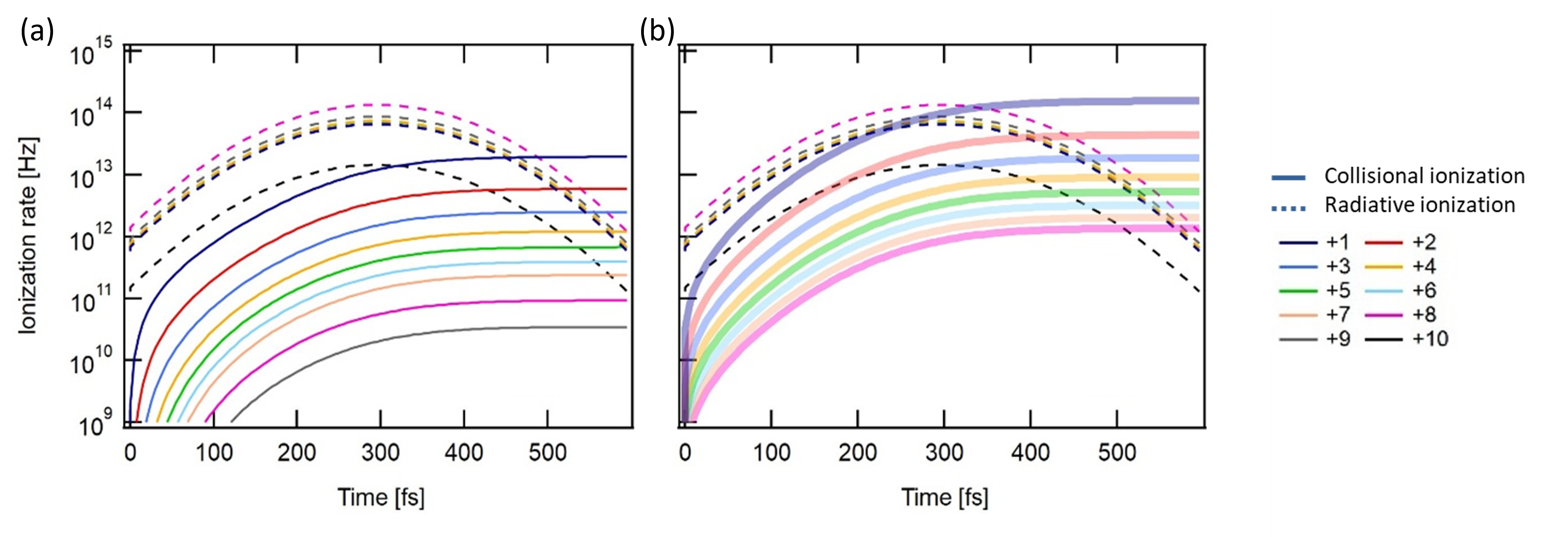}
    \caption{\label{rate_comp} Comparison of collisional ionization rate between without (a) and with non-equilibrium electron contributions during the XFEL pulse for charge states +1 to +10. The dotted line represents the photo-ionization rate for each ion, and it is consistent in both graphs. These plots are presented side-by-side to facilitate a comparison of collisional ionization rates. The XFEL pulse has the pulse duration of 230 $fs$ whose pulse center is on 300 $fs$.}
\end{figure*}

The thermalization time in neon plasma spans several picoseconds, leading to significant implications for the collisional ionization rate in the presence of non-equilibrium electrons. In high-density plasmas, the assumption of a Maxwellian electron distribution is valid due to rapid thermalization. However, neon plasma, with its lower density, experiences longer thermalization timescales, allowing energetic electrons to persist longer than in cases assuming instantaneous thermalization. Consequently, the collisional ionization rate is significantly enhanced when non-equilibrium electrons are considered.

Figure \ref{rate_comp} illustrates the collisional ionization rate over time, with the left panel assuming instantaneous thermalization and the right panel demonstrating the impact of non-equilibrium electrons on collisional ionization rates. In Fig. \ref{rate_comp}(a), we observe that, under the assumption of a Maxwellian electron distribution, the collisional ionization rate for most ions is smaller than the photoionization rate. However, the +1 ion, exhibiting the fastest crossover, is predominantly influenced by photoionization for approximately 300 femtoseconds. It is evident that photoionization plays a significant role in the ionization process during the XFEL-material interaction.

However, the presence of non-equilibrium electrons substantially enhances the ionization degree, resulting in a 1$\sim$2 orders of magnitude increase in the collisional ionization rate, shown in the Fig. \ref{rate_comp}(b). The effect is pronounced across all charge states, with lower charge states, such as +1 ions, experiencing a particularly significant impact. The collisional-ionization rate surpasses the photo-ionization rate during the middle of the simulation time frame, becoming the dominant process. Non-equilibrium electron effects enhance the collisional ionization rate by a factor of ten for the lower charge states in later stages of the process. Note that the non-equilibrium electrons alter the Auger and its inverse rates in this simulation. However, we have confirmed that their magnitudes are smaller by an order of magnitude than those of the collisional rates across all charge states, both with and without non-equilibrium electrons. This difference does not significantly impact the charge state distributions.

As a result, non-equilibrium electrons alter the pattern of ionization, making collisional ionization rates dominant from the peak of the pulse, while under thermalization without the consideration of non-equilibrium electrons, collisional ionization rates typically dominate over photo-ionization rates during the latter half of the process. The presence of non-equilibrium electrons proves to be crucial in understanding and modeling the collisional ionization dynamics in neon plasma.

\section{Conclusion}
The CR model characterizes physical processes in plasma based on electron temperature and density, emphasizing collisional processes for dense plasmas. However, its assumption of instantaneous electron thermalization has been shown not be accurate for low-density plasmas exposed to XFEL. To address this, a Boltzmann equation solver has been introduced to monitor the time evolution of the electron distribution function. This solver accounts for various types of collisions and atomic kinetic processes. By transforming the Boltzmann transport equation, the solver updates the electron distribution, highlighting non-equilibrium effects in XFEL-heated plasmas. Despite some complexities, the integration of the CR model and the Boltzmann solver improves the precision of plasma models, especially for low-density XFEL-plasma interactions.

To investigate the influence of non-equilibrium electrons on plasma dynamics, we studied the dynamics of ultrafast electron response in neon plasma exposed to intense XFEL X-ray pulses. The XFEL interaction with the low density ($n_e$ $\sim$ $10^{19}$ $cm^{-3}$) neon plasma requires the development of a new code that merges the collisional-radiative model with the Boltzmann solver. This code successfully tracks the EEDF's time evolution, revealing non-equilibrium electrons from photoionization and Auger decay. Comparison of simulation results with experimental data highlights discrepancies when non-equilibrium electrons are ignored. Non-equilibrium electrons substantially enhance the collisional ionization rate, surpassing the photoionization rate at certain simulation points. Considering non-equilibrium effects in neon plasma's interaction with XFEL pulses is crucial for accurately depicting collisional ionization dynamics.

In conclusion, we present a comprehensive study on the effect of non-equilibrium electrons in XFEL-heated plasma interactions. By introducing a Boltzmann equation solver into the CR model, we accurately represent the time evolution of the electron distribution function, revealing non-equilibrium electrons with longer lifetimes. The consideration of non-equilibrium effects is vital for accurately modeling XFEL-plasma interactions in low-density environments. Our combined approach of the CR model and the Boltzmann solver improves our understanding of collisional ionization dynamics and provides valuable insights for interpreting experimental results. The new code will serve as a powerful tool for researchers investigating plasma interactions with intense X-ray pulses in various low-density plasma scenarios.\\

This material is based upon work supported by the National Research Foundation of Korea (NRF-2019R1A2C2002864, RS-2023-00218180). This work also was performed under the auspices of the U.S. Department of Energy by Lawrence Livermore National Laboratory under Contract No. DE-AC52-07NA27344. We extend our deepest gratitude to Richard W. Lee and William L. Morgan, whose pioneering work two decades ago laid the foundational stones for our exploration into the impact of non-thermal electrons on NLTE kinetics.

\bibliography{ref.bib}

\end{document}